\documentclass[a4paper]{JHEP3}
\usepackage[dvips]{graphicx}

\usepackage{amsmath,amssymb}
\usepackage{cite}

\newcommand{\tk}{\Tilde k}
\newcommand{\tU}{\Tilde U}
\newcommand{\dd}{\delta(y-y_b)}

\newcommand{\m}{\mu}
\newcommand{\n}{\nu}
\newcommand{\la}{\lambda}
\newcommand{\La}{\Lambda}
\newcommand{\ro}{\rho}
\newcommand{\s}{\sigma}
\newcommand{\de}{\delta}
\newcommand{\al}{\alpha}
\newcommand{\be}{\beta}

\newcommand{\no}{\notag \\}

\title{Floating Black Hole in the Karch-Randall Model \\
and its Holographic Dual}

\author{
Norihiro Tanahashi${}^{a}$ and
Takahiro Tanaka${}^{b}$ \\
\llap{$^a$}Department of Physics, Kyoto University, 
Kyoto 606-8502, Japan \\
\llap{$^b$}Yukawa Institute for Theoretical Physics (YITP), \\
Kyoto University,Kyoto 606-8502, Japan \\
\email{
tanahashi@tap.scphys.kyoto-u.ac.jp,
tanaka@yukawa.kyoto-u.ac.jp}}

\abstract{
To investigate the holography in the Karch-Randall (KR) braneworld model, 
 we construct time-symmetric initial data of black holes floating in the bulk,
and compare it with its holographic dual, which is
described by
four-dimensional self-gravitating quantum field theory in 
asymptotically AdS$_4$ spacetime.
We also give a definition and an explicit formula
of mass in the KR model extending the definition by Abbott and Deser for 
asymptotically AdS spacetime.
We obtain supporting evidence for the holography in the KR model
such as good agreements of phase structures and characteristic values between the two theories,
and find clues that the Hawking-Page transition 
of the four-dimensional quantum theory in a microcanonical ensemble
is holographically dual to a transition in the bulk black hole configuration. 
}

\preprint{\arXivid{0910.5303}, KUNS-2238}	

\begin{document}

\section{Introduction}

The AdS/CFT correspondence~\cite{Mal} has been  attracting intensive attentions these years, 
and its extensions are discussed in wide settings.
In this paper, we focus on one of such an extension to the braneworld models.
The Randall-Sundrum (RS) model~\cite{RS} is a braneworld model that is composed of
infinitely extending AdS$_5$ bulk and a four-dimensional positive tension brane.
A remarkable feature of this model is that four-dimensional gravity is well reproduced on the brane
though the bulk spacetime is infinitely extending.
If we apply the AdS/CFT correspondence to the AdS bulk, we obtain a duality of new type
as follows~\cite{WittenComment,Verlinde,Gubser,HR}.
Since the AdS boundary is cut off by the brane in this model,
we may consider the CFT resides 
on the brane instead of the AdS boundary. It is believed that this CFT suffers an exotic UV cutoff by 
this introduction of the brane.
Adding to that, this deformed CFT couples to four-dimensional gravity on the brane 
in this setting.
Though this duality is not fully established, 
there are many pieces of circumstantial evidence for 
it and still attracting much 
interest~\cite{Shiromizu,Tanaka,Emparan,Kudoh,TanakaAdS,Fitz,Oriol,Tanahashi,Yoshino,Hubeny}.

\FIGURE[ht]{
\centering
\includegraphics[width=3.5cm,clip, angle=270]{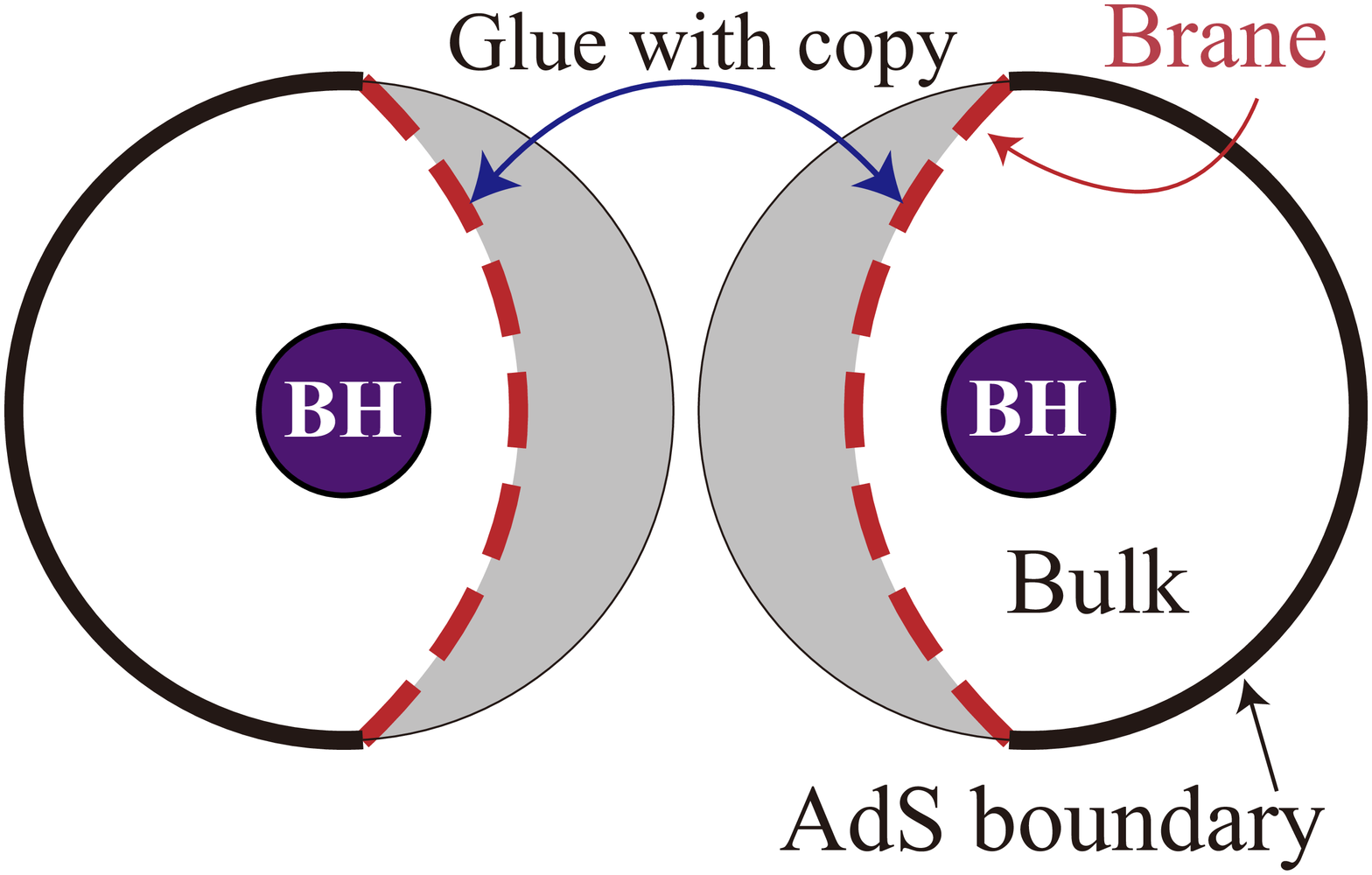}
\caption{
Schematic picture of the KR model, which 
is composed of $Z_2$-symmetric AdS$_5$ bulk regions and an AdS$_4$ brane.
We consider bulk black holes in this model.
}
\label{Fig:model_global}
}

An interesting extension of this duality is to modify the four-dimensional geometry on 
the brane from asymptotically flat to AdS.
This braneworld model with an asymptotically AdS brane is called Karch-Randall (KR) model~\cite{KR}, 
and its realization in the string theory is given in Ref.~\cite{KRstring}.
It is known that four-dimensional gravity is realized
for length scale shorter than $k^2/\tk^3$, where $\tk$ and $k$ are curvatures of the brane and 
the bulk respectively, 
though there are no normalizable zero mode and four-dimensional graviton is massive.
This massive graviton appears also 
in the four-dimensional gravity coupled to CFT
due to loop correction to the graviton propagator~\cite{P,Duff},
so the duality seems to hold even in this extended setting.
In the KR model, boundary of the bulk is a composite of the brane and the AdS boundary which share 
a common $S^2$ boundary, as is shown in Fig.~\ref{Fig:model_global}.
Then, the duality suggests that the bulk classical gravity corresponds to CFTs on those two boundaries.
Since the two boundaries can communicate via bulk in the five-dimensional perspective, 
the CFTs on the boundaries are suggested to communicate with each other through the common boundary~\cite{BR}.

In the holography of this type, the CFTs couple to gravity and it is in asymptotically AdS spacetime.
Then, the Hawking-Page transition may happen on the brane between thermal AdS phase and quantum-corrected
black hole phase if the duality holds even in the presence of strong gravity.
Based on this idea, 
it is conjectured that the transition in the five dimensions is 
 realized as a configurational transition between a black hole floating in the bulk 
and the one localized on the brane~\cite{TanakaFloat}.
A rough reasoning for it is as follows.
When the bulk black hole is sufficiently small, it can be trapped at the throat, 
at which the gravitational potential is minimized. This floating black hole will induce a 
localized effective energy density on the brane, which can be regarded as a star made of the CFT in 
hydrostatic equilibrium.
As we increase the bulk black hole size, it will touch the brane and become a brane-localized black hole.
An observer on the brane will regard this phenomenon as a transition from a CFT star to a
quantum-corrected black hole~\cite{TanakaFloat,Kashiyama}.
In this paper, we would like to test this conjecture by constructing explicit solutions in the 
four-dimensional and five-dimensional pictures%
\footnote{
By the way, Chamblin and Karch suggested that 
the five-dimensional counterpart of the four-dimensional Hawking-Page transition to be 
 a transition between themal AdS phase
and AdS black string phase~\cite{CK}.
Such a transition may happen if the system belongs to a canonical ensemble for fixed temperature, 
while it is impossible if the system is in a microcanonical ensemble for fixed energy
since generation of a bulk black string requires infinite energy.
In this paper, we consider phase transition in a microcanonical ensemble, which is more 
relevant to physical processes in a closed system, such as 
black object formation due to gravitational collapse on the brane.
}%
.

However, there are some problems to conduct comparison between the four-dimensional CFT stars and 
five-dimensional bulk black holes. 
Since the CFT stars are static objects, we should compare them with static solutions in 
five-dimensions, though we do not know such a static solutions so far.
Another problem is that we do not know much about self-gravitating CFT.
We cope with these difficulties by some approximations.
Firstly, we use five-dimensional time-symmetric initial data, which is a snapshot of a momentarily static spacetime, 
instead of static solutions.
This initial data will share some common properties with static solutions,
especially when its mass is minimized among the available set of initial data with the same entropy.
If the initial data set is complete, that initial data with minimum mass will be a static solution.
Next, we use radiation fluid approximation of the four-dimensional self-gravitating CFT, which is 
valid for high enough temperature~\cite{Kashiyama,Minwalla}.
We later see that this condition of high temperature is always satisfied in the regime we are 
interested in.
Using these approximations, we construct solutions numerically in four-dimensional and 
five-dimensional picture and test the conjecture of duality.
As a first step, we focus on the bulk floating black holes and the four-dimensional CFT stars, and 
study the duality between them.

One problem in the analysis is that 
there is no known good definition of mass so far in the KR model, which is 
necessary to know the thermodynamic properties of the five-dimensional system.
In this paper, we give an appropriate definition of mass in the KR model 
according to Abbott and Deser's definition of mass in an asymptotically AdS spacetime~\cite{AD}.
This is another feature of this work.

Organization of this paper is as follows.
We firstly illustrate how to construct a floating black hole initial data in Sec.~\ref{Sec:ID}.
In Sec.~\ref{Sec:explicit}, we propose a definition of mass in the KR model and give an explicit 
mass formula for initial data constructed in Sec.~\ref{Sec:ID}.
Using these tools, we analyze the initial data and summarize their basic properties in Sec~\ref{Sec:analysis}.
After that, we compare the five-dimensional bulk floating black holes with four-dimensional CFT 
stars in Sec.~\ref{Sec:CFTstar}. We briefly summarize how to construct CFT star solutions, and 
then we compare them with the initial data.
We find good agreements of phase structures and characteristic values between them.
Finally, we give concluding remarks in Sec.~\ref{Sec:summary}.

\section{Floating black hole initial data}
\label{Sec:ID}

In this section, we illustrate the construction method of time-symmetric initial data of floating black holes in 
the KR model%
\footnote{%
Our construction method is partially based on that used in Ref.~\cite{Creek}.
}. 
An initial data is a set of spatial geometry and its time derivatives which solves
the Hamiltonian constraint and momentum constraints in 
the Einstein equations.
In the case that the spacetime is time-symmetric about a $t=\text{constant}$ surface, the momentum 
constraints are trivially satisfied. Thus we have to solve only the Hamiltonian constraint in this case.

To construct an initial data of the KR model, we have to solve the five-dimensional Hamiltonian 
constraint in the bulk and the four-dimensional one on the brane
simultaneously.
In this work, we solve those equations by taking a well-known static solution in the bulk 
and embedding the brane that satisfies the constraints.

In the bulk, we take the AdS-Schwarzschild solution:
\begin{equation}
 ds^2 = -U dt^2 +\frac{dr^2}{U} + r^2
\left(
d\chi^2+\sin^2\chi\, d\Omega_\text{II}
\right)~~,
\qquad
U=1+k^2r^2-\frac{\mu}{r^2}~~,
\label{global}
\end{equation}
where $k$ is the five-dimensional bulk curvature, which is related to the five-dimensional 
cosmological constant as $\Lambda_\text{5}=-6k^2$.
$\mu$ is the mass parameter of the black hole, and
 $d\Omega_\text{II}$ is the metric of a unit two-sphere.
The black hole horizon is at $r_g$, the largest root of the equation $U(r)=0$.

\FIGURE[ht]{
\centering
\includegraphics[width=3.8cm,clip]{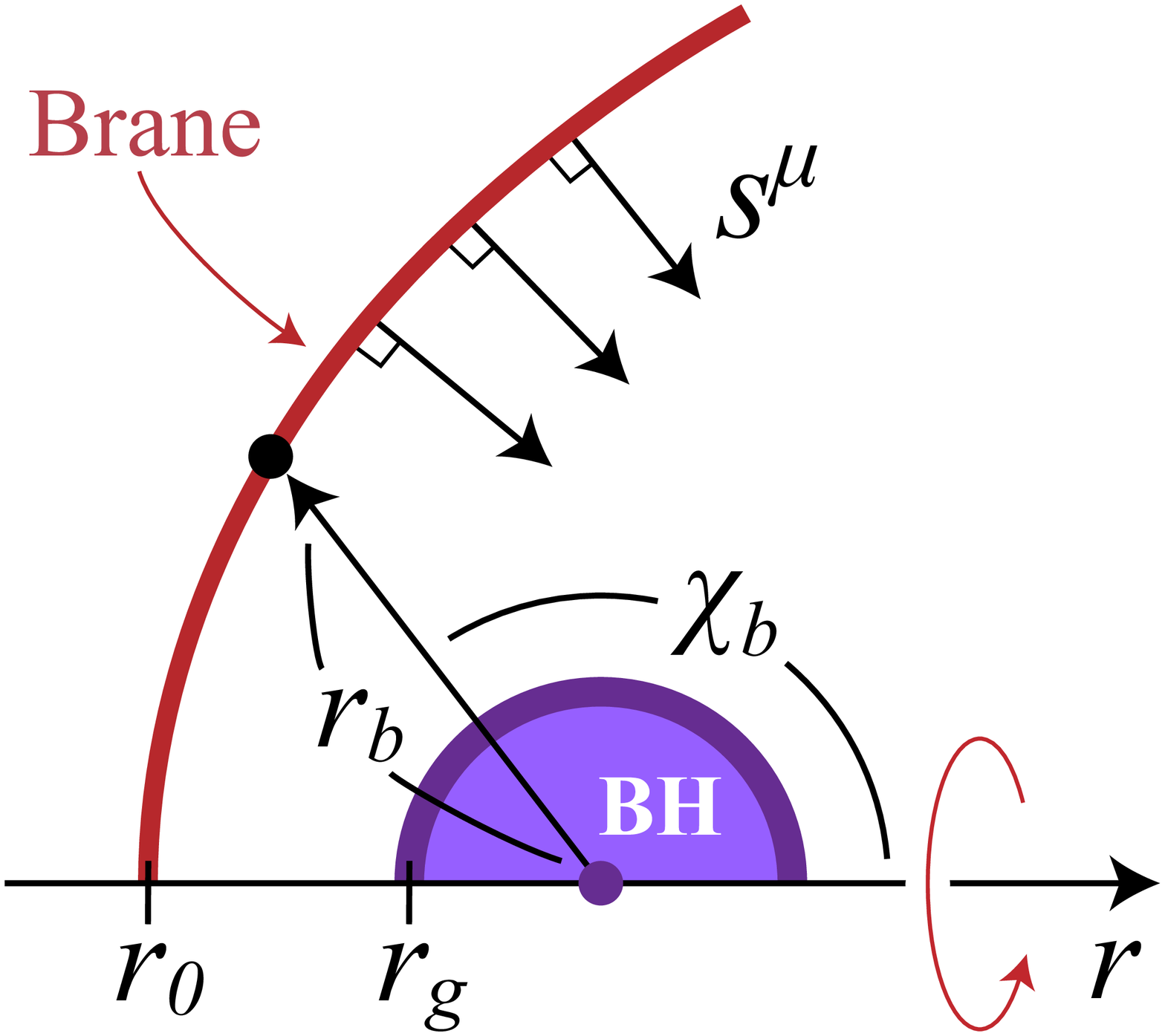}
\caption{
Configuration of the brane and the black hole in an
initial data in the global coordinates.}
\label{Fig:ID}
}

Taking this AdS-Schwarzschild black hole spacetime as a background, we determine the brane 
trajectory on it solving the four-dimensional Hamiltonian constraint.
Let us introduce four-dimensional induced metric on the brane $\gamma_{ab}$ and unit normal 
of the brane $s_\mu$ that extends from the brane into the bulk. 
Here the Latin indices ($a,b,\dots$) run through only the four-dimensional 
coordinates parallel to the brane. The extrinsic curvature of the brane is defined to be
$K_{ab}=-\gamma_a{}^\mu \gamma_b{}^\nu \nabla_\mu s_\nu$. 

In the KR model, the five-dimensional energy-momentum tensor along the brane $T_{ab}$ is given by
\begin{equation}
 T_{ab}=\sigma \gamma_{ab}~~,
\qquad
\sigma\equiv \frac{3\sqrt{k^2-\tk^2}}{4\pi G_5}~~,
\end{equation}
and is localized on a four-dimensional surface in the bulk.
$\sigma$ is the brane tension and $\tk$ is the four-dimensional curvature that is 
related to the four-dimensional cosmological constant as $\Lambda_\text{4}=-3\tk^2$,
and $G_5$ is the five-dimensional gravitational constant.
This model reduces to the RS model with a flat brane by setting $\tk=0$.
In this work, we focus on the regime that $k\gg\tk$, i.e.~the deformation from the RS model is small.

Using these quantities, Israel's junction conditions at the brane are written as
\begin{equation}
 K_{ab}-K\gamma_{ab}=\frac{1}{2}\cdot 8\pi G_5 T_{ab}~~,
\label{HC}
\end{equation}
where we used the $Z_2$-symmetry about the brane.
These equations are the 
five-dimensional Einstein equations integrated across the brane, and
their time-time and time-space components are 
the Hamiltonian constraint and the momentum constraints, respectively.
Let us assume here the time-symmetry about the $t=\text{constant}$ surface we are focusing on. Then, 
the momentum constraints are trivially satisfied, and we only have to solve the Hamiltonian 
constraint, which can be rewritten as 
\begin{equation}
 D_i s^i=-3\sqrt{k^2-\tk^2}~~,
\label{HC2}
\end{equation}
where $D$ is the covariant derivative with respect to $\gamma_{ab}$ and the index $i$ runs through 
only the three-dimensional spatial coordinates on the brane.

If we assume the $O(3)$ symmetry of the brane, its trajectory in the bulk can be parametrized as 
$(r,\chi)=\left(r_b(R), \chi_b(R)\right)$, where $R$ is the proper length along 
the brane measured from the axis of the $O(3)$ rotational symmetry.
Under this parametrization the Hamiltonian constraint~(\ref{HC2}) becomes a set of second order 
differential equations about $r_b$ and $\chi_b$. Supplying the regularity condition on the axis, 
those equations can readily be integrated.
Once the brane trajectory is determined, an initial data is obtained by cutting the bulk spacetime 
along the brane, taking the part of the bulk that contains the black hole, and gluing 
that bulk with its copy along the brane.
The resultant initial data possesses 
two copies of bulk regions separated by a four-dimensional asymptotically AdS brane,
and each bulk region encompasses a floating black hole.

In the above construction procedure,
we can set freely the mass parameter $\mu$ of the background AdS-Schwarzschild spacetime and the brane's 
starting point $r_b(0)\equiv r_0$. Thus, the initial data constructed by the above method constitute a two-parameter family.
By the way, if we make the brane starting point $r_0$ too close to the bulk black hole, the brane 
trajectory directly falls into the black hole and an asymptotically AdS brane is not realized. 
Examining Eq.~(\ref{HC2}), we find the lower bound on $r_0$, $r_\text{min}$, to realize 
an asymptotically AdS brane to be the largest root of the equation $U(r_\text{min})=r_\text{min}^2(k^2-\tk^2)$.

In some initial data, typically that with small $\mu$ and $r_0$, an apparent horizon appears that 
encloses the bulk black hole and crosses perpendicularly the brane. 
Apparent horizon appearance implies the existence of an event horizon outside of it. 
The black holes in such initial data are not floating in the bulk but are localized on the brane.
We neglect such initial data and focus only on 
those with floating black holes in this work.

Before closing this section, 
let us summarize the series solution of the brane trajectory in the asymptotic region for the later use.
The brane trajectory can be parametrized as $\chi=\chi_b(r)$. Expanding this expression in the 
asymptotic region as 
\begin{equation}
\chi_b(r)=\sum_{i=0}\chi_i (kr)^{-i}
\label{series}
\end{equation}
and substituting it into the Hamiltonian 
constraint~(\ref{HC2}), we find that 
$\chi_0$ is a free parameter and
$\chi_1$ is determined by $k$ and  $\tk$ 
as $\chi_1=\{(k/\tk)^2-1\}^{1/2}$. 
The coefficients $\chi_2$ and $\chi_3$ are functions of $\chi_0$ and $\chi_1$. 
The equation for $\chi_4$ is automatically satisfied, and 
the remaining coefficients $\chi_{i>4}$ are determined by $\chi_0$ and $\chi_4$.
Thus, 
the series solution in the asymptotic region constitutes a two-parameter family whose free parameters 
are $\chi_0$ and $\chi_4$.
We see later that we can freely change the parameter $\chi_0$ by the Lorentz transformation, and that
$\chi_4$ is related to the mass contained in the initial data.

\section{The Abbott-Deser mass of initial data}
\label{Sec:explicit}

To analyze the thermodynamic property of black holes in the KR model,
we have to know about its mass.
However, there is no known definition of mass in the KR model so far and it is non-trivial how to 
define it.
In this section, we show that we can define conserved mass by generalizing the definition of mass in 
asymptotically AdS spacetime by Abbott and Deser.
We also illustrate in detail
how to calculate this Abbott-Deser mass of initial data constructed in Sec.~\ref{Sec:ID}.
The Abbott-Deser mass we derive in Sec.~\ref{Sec:mass}
is given by a surface integral of metric perturbation 
in the coordinates that are Gaussian normal (GN) near the brane.
We refer to it simply as GN coordinates.
Since the initial data is constructed in the AdS global coordinates,
we have to transform the coordinates into the GN coordinates to calculate the mass.
We summarize the coordinate transformation procedure in Sec.~\ref{Sec:coord},
and study the metric perturbation due to the bulk black hole in Sec.~\ref{Sec:Sch}.
Finally, we give an explicit formula of the Abbott-Deser mass of the initial data in Sec.~\ref{Sec:formula}.

\subsection{The Abbott-Deser mass in the KR model}
\label{Sec:mass}

In this section, 
we give a definition of mass in the KR model extending the definition by Abbott and Deser for asymptotically
AdS spacetime.
A naive extension results in an apparent violation of the mass conservation law due to the 
brane's contribution.
We show that this problem can be circumvented at least if we introduce the GN coordinates with respect to the 
brane, and that conserved mass can be appropriately defined.

The five-dimensional Einstein equations in the KR model are given by
\begin{equation}
 G_{\mu\nu}\equiv
R_{\mu\nu}-\frac{1}{2}Rg_{\mu\nu}+\Lambda_\text{5} g_{\mu\nu}
+
\frac{\sigma\dd }{\sqrt{g_{yy}}}
\gamma_{\mu\nu}=0~~,
\label{Eeq}
\end{equation}
where the last term is the contribution of the brane. The coordinate $y$ in the delta function is 
the one perpendicular to the brane in the GN coordinates in which the brane is at
$y=\text{constant}$ surface at $y=y_b$ and the metric is taken to be 
$g_{y\mu}=0$ for $\nu\neq y$ on the brane.

Let us consider a background solution $\bar g_{\mu\nu}$ and 
a perturbed solution $g_{\mu\nu}$ given by $ g_{\mu\nu} = {\Bar g}_{\mu\nu} + h_{\mu\nu}$.
We assume that both ${\Bar g}_{\mu\nu}$ and $g_{\mu\nu}$ solve the Einstein equations~(\ref{Eeq}),
and that ${\Bar g}_{\mu\nu}$ is static and possesses a timelike Killing vector $\Bar \xi^\mu$.
We define $\delta G_{\mu\nu}$ to be $\mathcal{O}(h)$ part of the Einstein tensor.
Then, this $\delta G_{\mu\nu}$ satisfies the first order Bianchi identity $\delta G_{\mu\nu}{}^{;\nu}=0$,
where the semicolon is the background covariant derivative with respect to $\Bar g_{\m\n}$,
by virtue of the Einstein equations for ${\Bar g}_{\mu\nu}$.

Using these equations, conserved energy can be defined as follows.
Firstly, $\delta G_{\mu\nu}$ can be rewritten as 
\begin{align}
 \delta G_{\mu\nu}=&
\phantom{+}
K_{\mu\alpha\nu\beta}{}^{;\be\al}
+\frac{1}{2}\left(
\Bar R_{\n\rho}H^\rho{}_\m + \Bar R_{\la\n\m}{}^\rho H^{\la}{}_{\rho}\right)
\no
&
+\frac{1}{2}h^{\la\rho}\Bar R_{\la\rho}g_{\m\n}
-\frac{1}{2}\Bar R h_{\m\n}+\La h_{\m\n}
+\de\left\{
\frac{\s\dd }{\sqrt{g_{yy}}}
\gamma_{\m\n}\right\},
\label{deltaG}
\end{align}
where the $K_{\m\al\n\be}$ and $H_{\m\n}$ are defined as
\begin{equation}
 K_{\m\al\n\be}\equiv
\frac{1}{2}\left(
\Bar g_{\m\be}H_{\n\al}+\Bar g_{\n\al}H_{\m\be}
-\Bar g_{\m\n}H_{\al\be}-\Bar g_{\al\be}H_{\m\n}
\right)~,
\qquad
H_{\m\n}\equiv h_{\m\n}-\frac{1}{2}h\Bar g_{\m\n}~,
\end{equation}
and indices are raised by the background metric ${\Bar g}_{\m\n}$.
$\de\left\{\s \dd \gamma_{\m\n}/\sqrt{g_{yy}}\right\}$ 
is $\mathcal{O}(h)$ part of $\sigma \dd \gamma_{\mu\nu}/\sqrt{g_{yy}}$.
Contracting Eq.~(\ref{deltaG}) with $\Bar\xi^\mu$
and using the background Einstein equations ${\Bar G}_{\m\n}=0$, we obtain
\begin{align}
 \delta G_{\mu\nu}\Bar\xi^\nu
=&
\left(
K_{\m\al\n\be}{}^{;\be}\Bar\xi^\n - K_{\m\be\n\al}\Bar\xi^{\n;\be}
\right)^{;\al}
\no&
+\frac{\s\dd}{2\sqrt{\Bar g_{yy}}}\left(
-
2\Bar \gamma^{\la\n}h_{\m\n}\Bar\xi_\la
+
h\Bar\gamma_\m{}^\la\Bar\xi_\la
-
\Bar\gamma_{\ro\la}h^{\ro\la}\Bar\xi_\m
\right)
+\de\left\{
\frac{\s\dd}{\sqrt{g_{yy}}}
\gamma_{\m\n} \right\}\Bar\xi^\n~.
\label{dGxi}
\end{align}
Then, we define the Abbott-Deser mass $M$ as
\begin{equation}
 M \equiv \frac{1}{8\pi G_5}\int_{\Sigma}d^4x \sqrt{-{\Bar g}} \;\delta G^{\tau\nu}\Bar\xi_\nu~~,
\label{mass}
\end{equation}
where $\tau$ is the time coordinate of the GN coordinates, and
the integral is taken over a large four-dimensional volume $\Sigma$.
When there are no brane, i.e.~$\sigma=0$, Eq.~(\ref{dGxi}) becomes a total divergence,
and we can transform Eq.~(\ref{mass}) into  a surface integral 
over three-dimensional spacelike surface $\partial\Sigma$.
It implies the conservation of $M$.

We show below that this conservation law holds even when the brane exists, 
at least in the GN coordinates.
In our setting, the background spacetime including the brane is assumed to be static
and the coordinates are taken to be in the GN coordinates, 
i.e.~$\Bar\xi^\mu=0=\Bar\gamma_{0\mu}$ for $\mu\neq t$ and 
$\Bar g_{y\m}=0$ for $\m\neq y$.
Then, the terms in Eq.~(\ref{dGxi})
other than the total divergence term reduce to
\begin{multline}
\frac{\sigma\dd}{2\sqrt{\Bar g_{yy}}}
\!
\left(
-2\Bar\gamma^{\tau\n}h_{\tau\n}
+ h-\Bar\gamma_{\ro\s}h^{\ro\s}
\right)
\!
\Bar\xi_\tau
+\de\left\{ 
\frac{\s\dd}{\sqrt{g_{yy}}}
\gamma_{\tau\tau}\right\}
\!
\Bar\xi^\tau
\\
=
\frac{\s\dd}{2\sqrt{\Bar g_{yy}}}
\left(
-2h^\tau_{~\tau}
+h^y_{~y}
\right)\Bar\xi_\tau
+\de\left\{ 
\frac{\s\dd}{\sqrt{g_{yy}}}
\gamma_{\tau\tau}\right\}\Bar \xi^\tau~.
\label{otherterms}
\end{multline}
Let us assume here that the perturbed metric $g_{\m\n}$ is 
written in the GN coordinates, i.e.~$h_{y\m}=0$ for $\m\neq y$.
Under this condition, the second term in the right-hand side of Eq.~(\ref{otherterms}) becomes
$\s\dd (h^\tau_{~\tau} - h^y_{~y}/2)\Bar\xi_\tau$~, and 
it cancels the first term.
Then, 
the Abbott-Deser mass~(\ref{mass}) 
becomes a surface integral as
\begin{align}
 M &= \frac{1}{8\pi G_5}\int_{\Sigma}d^4x \sqrt{-{\Bar g}} 
\left(
K^{\tau\al\n\be}{}_{;\be}\Bar\xi_\n - K^{\tau\be\n\al}\Bar\xi_{\n;\be}
\right)_{;\al}
\no
&= \frac{1}{8\pi G_5}\oint_{\partial\Sigma} dS_\al \sqrt{-\Bar g}\left[
K^{\tau \al\n\be}{}_{;\be}\Bar \xi_\n-K^{\tau \be\n \al}\Bar \xi_{\n;\be}
\right]~~,
\label{mass2}
\end{align}
and thus it is conserved.
A proof of the Abbott-Deser mass conservation in a general coordinate system 
is probably possible, but we do not investigate it here.

\subsection{Coordinate transformation of the background geometry}
\label{Sec:coord}

The goal of the following subsections is to calculate the Abbott-Deser mass of the initial data we constructed.
The expression of the Abbott-Deser mass we defined is given in terms of the metric perturbation 
in the GN coordinates, while the initial data is constructed in the AdS global coordinates~(\ref{global}). 
Thus, we have to transform the coordinates first for the mass calculation.
We consider the transformation of the background spacetime with no bulk black hole in this 
subsection, and after that we take into account the geometry deformation due to the bulk black hole 
in the next subsection.

When there is no bulk black hole, the metric of the KR model 
in the GN coordinates is given by
\begin{equation}
 ds^2=dy^2+\frac{\tk^2}{k^2}\cosh^2\left(ky\right)
\left(
-\tU d\tau^2+\frac{1}{\tU}d\la^2+\la^2d\Omega_\text{II}
\right)~, \qquad
\tU=1+\tk^2 \la^2~~,
\label{GN}
\end{equation}
where the brane location is determined by $(\tk/k)\cosh(ky_b)=1$~.
We can find the coordinate transformation 
from the global coordinates~(\ref{global}) to these GN coordinates~(\ref{GN})
by comparing the metric components, which is summarized as
\begin{equation}
 kr\cos\chi=\sinh\left(ky\right)~, \quad
kr\sin\chi=\tk \la\cosh\left(ky\right)~, \quad
U =\tU \cosh^2\left(ky\right)~, \quad
kt =\tk \tau~.
\label{trsf}
\end{equation}
By this transformation, the brane trajectory in the GN coordinates is mapped to that in the global 
coordinates whose asymptotic behavior for $r\to\infty$ is given by $\chi_0=\pi/2$, as we can see from 
the first equation of~(\ref{trsf}).

However,  $\chi_0$ does not become $\pi/2$ for a general initial data.
Hence, we first perform a transformation in the global coordinates so that $\chi_0$ becomes $\pi/2$.
Such a transformation is given by
\begin{equation}
 r'\sin\chi'=r\sin\chi~~,
\qquad
r'\cos\chi'=\gamma\left(r\cos\chi-v\sqrt{r^2+k^{-2}}  \right)~~,
\label{Lorentz}
\end{equation}
where $v=\cos\chi_0$ and $\gamma=(1-v^2)^{-1/2}$. This transformation is the Lorentz 
transformation of the higher-dimensional embedding of AdS %
\footnote{
AdS$_5$ spacetime~(\ref{global}) can be embedded into six-dimensional pseudo-Euclidean space 
$E^{4,2}$ by
\begin{gather}
 z_0=\sqrt{r^2+k^{-2}}\sin(kt)~, \quad
z_1=\sqrt{r^2+k^{-2}}\cos(kt)~, \quad
z_2=r\sin\chi\sin\theta\cos\phi~,
\no
z_3=r\sin\chi\sin\theta\sin\phi~, \quad
z_4=r\sin\chi\cos\theta~, \quad
z_5=r\cos\chi
\notag
\end{gather}
as a hyperboloid 
$-z_0{}^2-z_1{}^2+\sum_{i=2}^6 z_i{}^2=-k^{-2}$. 
The Lorentz transformation in six dimensions given by
\begin{equation}
z'_5=\gamma\left(z_5-v z_1\right)~, \quad 
z'_1=\gamma\left(z_1-vz_5\right)
\notag
\end{equation}
maps the hyperboloid into itself, and its restriction to the $t=0$ surface
generates the transformation~(\ref{Lorentz}).
This restricted transformation is sufficient for our analysis since the explicit 
calculation shows that the integrand of Eq.~(\ref{mass2}) does not contain neither 
the time-time component of the metric perturbation 
nor derivatives of the other components with respect to the time coordinate.
}.
The metric~(\ref{global}) is kept invariant by it,
while the brane is translated by this transformation from the original position 
$\chi=\chi_0+\mathcal{O}(r^{-1})$ to the desired position $\chi'=\pi/2+\mathcal{O}(r'^{-1})$.
By applying two transformations~(\ref{Lorentz}) and (\ref{trsf}) successively, 
we can convert an initial data with no bulk black hole into the KR model written in the GN coordinates.
If we trace back this transformation from the metric~(\ref{GN}), we find that the asymptotic 
trajectory of a pure AdS brane in the global coordinates is given by
\begin{equation}
\chi_4=-\frac{k^2\cos\chi_0\left\{ \left(3+2\cos^2\chi_0\right)k^2-4\tk^2  \right\}}{8\tk^4\sin^3\chi_0}
\label{chi4}
\end{equation}
for general $\chi_0$.

\subsection{Perturbation by the black hole}
\label{Sec:Sch}
In this subsection, we consider the case that a bulk black hole exists.
In this case, the bulk geometry and the brane trajectory are deformed and the coordinate 
transformation~(\ref{trsf}) and (\ref{Lorentz}) does not give a metric in the GN coordinates.
To be more precise, the brane trajectory in the asymptotic region acquires deviation of $\mathcal{O}(r^{-4})$
due to the bulk black hole, and it results in violation of  the GN coordinate condition: 
$h_{y\mu}=0$ for $\mu\neq y$ on the brane.
To restore the GN coordinate condition, we have to apply an additional gauge transformation. 
As a result, the total metric perturbation is given by a sum of the contribution from the background metric 
deformation and that from the additional gauge transformation.

Let us formulate the deviation due to the bulk black hole explicitly.
When the bulk black hole exists,
a brane trajectory in the global coordinates is deformed as
\begin{equation}
 \chi(r)=\chi^{BG}(r) + \frac{\delta\chi_4}{\left(kr\right)^4} + \mathcal{O}\left( r^{-5} \right)~,
\end{equation}
where $\chi=\chi^\text{BG}(r)$ represents a brane trajectory of a pure AdS brane.
The order of the brane trajectory deviation, $r^{-4}$, is known from 
the form of the series solution for $r\to\infty$ (see Eq.~(\ref{series}) and below).
Applying the background coordinate transformation~(\ref{trsf}) and (\ref{Lorentz}), 
we obtain the following brane trajectory:
\begin{equation}
y=y_0+\frac{\delta y_3}{\big(\tk \la\big)^3}+\mathcal{O}\left( \la^{-4}  \right)~,
\qquad
\delta y_3 = -\frac{\tk^4\sin^3\chi_0}{k^5}\delta\chi_4~~.
\end{equation}
To obtain a metric that satisfies the GN condition, we have to eliminate this deviation $\delta y_3$.
It is realized by an additional gauge transformation defined as 
\begin{equation}
 x'^\mu=x^\mu+\zeta^\mu~~,
\qquad
\left( \zeta^y, \zeta^\la \right)=\left(
-\frac{\delta y_3 F(y)}{\big( \tk \la \big)^3} , \frac{G(y)}{\big( \tk \la  \big)^2}
\right)~~,
\end{equation}
where $F(y)$ and $G(y)$ are arbitrary functions that vanish for $y\to\infty$ and satisfy $F(y_b)=1$.
To satisfy the GN coordinate condition, we require later an additional condition on $G(y)$.

In the resultant GN coordinates, the metric perturbation from the background~(\ref{GN}) is given by
$ h_{\m\n}= h^\text{BH}_{\m\n} + h^\zeta_{\m\n}$,
where $h^\text{BH}_{\m\n}$ comes from the metric perturbation due to the AdS-Schwarzschild black hole, 
that is $h^\text{global}_{rr}=U^{-1}-(1+k^2r^2)^{-1}$ in the global coordinates.
This perturbation is transformed to that in the GN coordinates as
$h^\text{BH}_{\m\n}=(\partial r/\partial x^\m)(\partial r/\partial x^\n)h^\text{global}_{rr}$, where
$\m$ and $\n$ are $y$ or $\la$.
$h^\zeta_{\m\n}=-2\zeta_{(\m;\n)}$ is the contribution from the gauge transformation $\zeta^\m$.
In the asymptotic region $\la\to\infty$, 
the leading term with respect to $\la$ of 
the $(y,\la)$ component of each metric perturbation becomes
\begin{equation}
 h^\text{BH}_{y\la} = 
\frac{\m k \left(1-v^2\right)^2 \sinh\left(ky\right) }{\tk^4\la^5\cosh^5\left(ky\right)}~~,
\qquad
h^\zeta_{y\la}= 
-\frac{3k^2F(y_b)\delta y_3+\tk G'(y_b)\cosh^2\left(y_b\right)}{k^2\tk^3\la^4}~~.
\end{equation}
Then, we find that the GN coordinate condition $0=h_{y\la}=h^\text{BH}_{\m\n}+h^\zeta_{\m\n}$ is 
satisfied for $G'(y_b) = -3\tk \delta y_3$, where we used the relation $(\tk/k)\cosh\left(ky_b\right)=1$
and the  condition $F(y_b)=1$.
This formula for $G'(y_b)$ is correct only in the asymptotic region $\la\to \infty$, but it is sufficient to 
calculate the Abbott-Deser mass as we see in the next subsection.

\subsection{Explicit formula of mass}
\label{Sec:formula}

In this subsection, we give an explicit formula of the Abbott-Deser mass (\ref{mass2})
for floating black hole initial data using the formulae obtained in the previous subsections.

Some analysis shows that 
the integrand of Eq.~(\ref{mass2}) is a function of 
the spatial components of the metric perturbation $h_{\m\n}$, which is calculated in Sec.~\ref{Sec:Sch}.
We normalize the timelike Killing to be $\Bar\xi^\tau=-1$, in other words we measure the mass with 
respect to the time coordinate on the brane.
To perform an explicit integration, we take the boundary $\partial \Sigma$ to be a cylinder of radius $\la_B$
that is truncated at $y=y_B$ as shown in Fig.~\ref{Fig:surf}.
 After the integration, we take the limit of $\la_B$ and  $y_B\to\infty$.

\FIGURE[ht]{
\centering
 \includegraphics[width=4cm,clip]{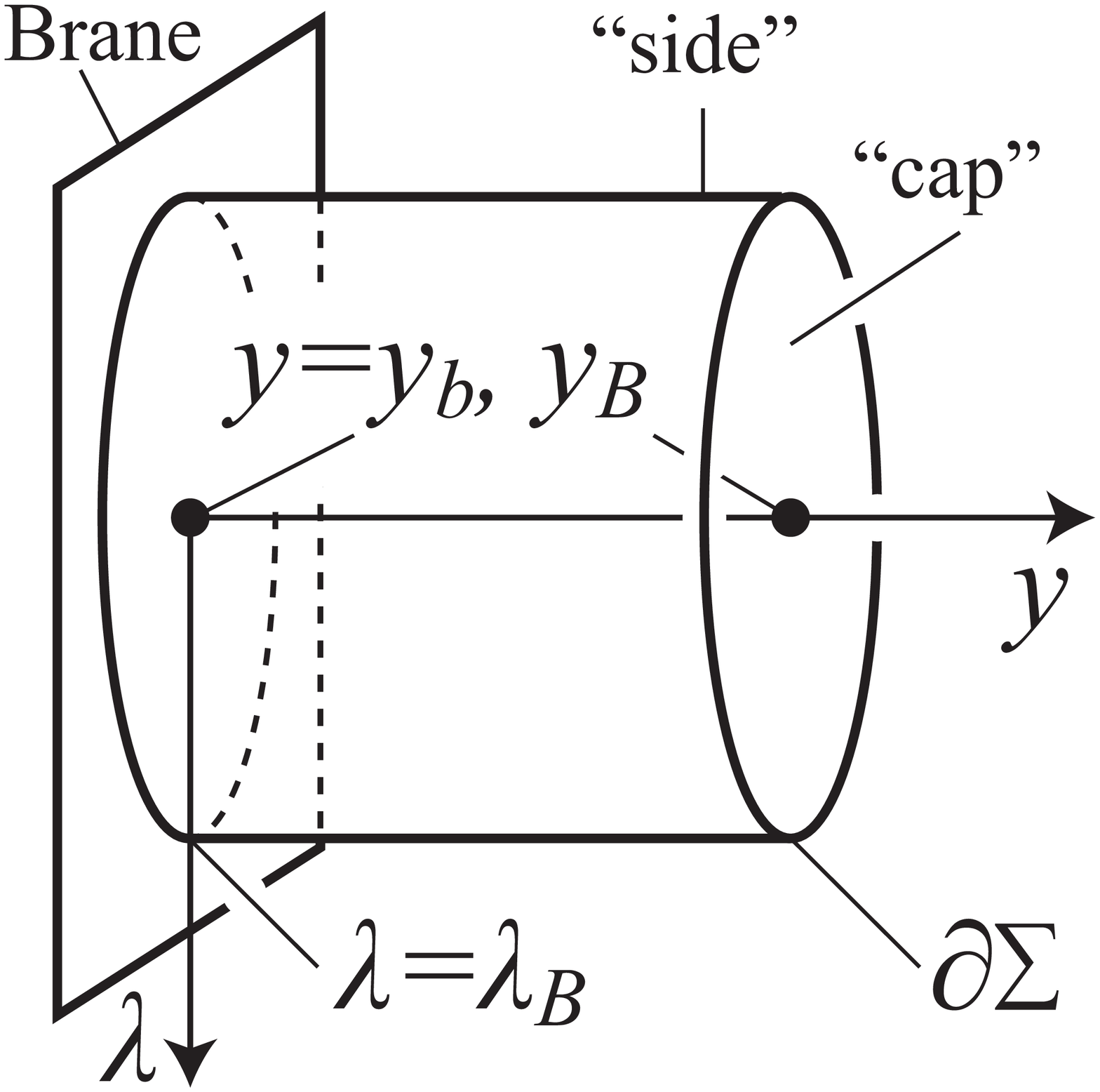}
\caption{
Integration surface.
}
\label{Fig:surf}
}

Firstly, the integration on the ``side'' of $\partial \Sigma$, that is 
the cylindrical surface $\la=\la_B$, is given as
\begin{align}
&~
2\times\frac{1}{8\pi G_5}\oint_\text{side} 
\!\!\!\!\!\!
dy d\theta d\phi \sqrt{-\Bar g}
\left[
K^{\tau \la\n\be}{}_{;\be}\Bar \xi_\n-K^{\tau \be\n \la}\Bar\xi_{\n;\be}
\right]
\no
&=\frac{\tk^4 \la_B^2}{k^4 G_5}
\int^{y_B}_{y_b}
\!\!\!
dy\cosh^4\left(ky\right)
\left[
K^{\tau \la\n\be}{}_{;\be}\xi_\n-K^{\tau \be\n \la}\xi_{\n;\be}
\right]
\bigg|_{\la=\la_B}
\;.
\label{intSide}
\end{align}
The factor two in the left-hand side
is multiplied since the same contribution comes from the both 
sides of the brane.
For the perturbation due to the black hole $h^\text{BH}_{\m\n}$, 
the integrand of Eq.~(\ref{intSide}) becomes $\mathcal{O}\left(\la_B^{-1}\right)$ and thus the 
integral vanishes in the limit $\la_B\to\infty$.
On the other hand,
the contribution from the gauge perturbation $h^\zeta_{\m\n}$ becomes
\begin{align}
&\lim_{\la_B, y_B\to\infty}
\frac{\tk^4 \la_B^2}{k^4 G_5}
\int^{y_B}_{y_b}
\!\!\!
 dy 
\cosh^4\left(ky\right)
\left[
K^{\tau \la\n\be}{}_{;\be}\xi_\n-K^{\tau \be\n \la}\xi_{\n;\be}
\right]
\no
&~~~~
=
-\frac{\tk^3}{2k^4G_5}
\left[
\cosh^2(ky)\left(
\frac{G'\cosh^2(ky)}{\tk}
-
%
%
\frac{5k^2\delta y_3 F}{\tk^2}
\right)
\right]^{y=\infty}_{y=y_b}
=
-\frac{4\delta y_3}{\tk G_5}
=\frac{4\tk^3\sin^3\chi_0}{k^5 G_5}\delta\chi_4
~,
\label{sideGauge}
\end{align}
where we used the boundary conditions for $F$ and $G'$ as well as $(\tk/k)\cosh\left(ky_b\right)=1$.

Next, the integration on the ``caps'' of the cylinder, that is the surfaces $y=\pm y_B$, is given as
\begin{align}
&~
2\times\frac{1}{8\pi G_5}\oint_\text{cap} 
\!\!\!\!\!
d\la d\theta d\phi 
\sqrt{-\Bar g}
\left[
K^{\tau y\n\be}{}_{;\be}\Bar \xi_\n-K^{\tau \be\n y}\Bar \xi_{\n;\be}
\right]
\no
&=
\frac{\tk^4 \cosh^4\left(ky_B\right)}{k^4 G_5}
\int_{0}^{\la_B}
\!\!\!
d\la\,\la^2
\left[
K^{\tau y\n\be}{}_{;\be}\Bar \xi_\n-K^{\tau \be\n y}\Bar\xi_{\n;\be}
\right]
\bigg|_{y=y_B}
\;.
\label{capIntegral}
\end{align}
$h^\text{BH}_{\m\n}$ gives a nonzero contribution to this integral as
\begin{align}
&\lim_{\la_B, y_B\to\infty}
\frac{\tk^4 \cosh^4\left(ky_B\right)}{k^4 G_5}
\int_{0}^{\la_B}
\!\!\!
d\la
\, \la^2
\left[
K^{\tau y\n\be}{}_{;\be}\xi_\n-K^{\tau \be\n y}\xi_{\n;\be}
\right]
\no
&~~~=
\int_{0}^{\infty}
\!\!\!
d\la
\,
\frac{3\mu\tk^4 \left(1-v^2\right)^2 \la^2}{2kG_5\big( \tU^{1/2}+v \big)^4}
=
\frac{\m\tk}{4kG_5\sin\chi_0}
\left\{
3\chi_0+\sin\chi_0\cos\chi_0\left(
2\cos^2\chi_0-5
\right)
\right\}\;,
\label{capSch}
\end{align}
while $h^\zeta_{\m\n}$ gives no contribution since $\zeta^\m$ vanishes for $y\to\infty$.

After all, we obtain an expression of 
the Abbott-Deser mass as 
\begin{equation}
 M=
\frac{1}{k^2 G_5}
\left[
\frac{4\tk^3\sin^3\chi_0}{k^3}\,\delta\chi_4
+
\frac{\m k\tk}{4\sin\chi_0}
\left\{
3\chi_0+\sin\chi_0\cos\chi_0\left(
2\cos^2\chi_0-5
\right)
\right\}
\right]~~,
\label{massFinal}
\end{equation}
which is totally expressed in terms of the quantities in the global coordinates.
Once we construct an initial data numerically,
its mass can be calculated by reading out 
 $\chi_0$ and $\delta \chi_4$ from the brane trajectory and plugging those into the above formula.

Let us comment on the tensionless limit of the mass formula, that is the limit to make $\chi_0\to\pi/2$
and $k\to\tk$ simultaneously. 
In this limit, the five-dimensional system reduces to an AdS-Schwarzschild black hole with a 
tensionless brane on its equator plane, which do not contribute to the mass at all.
This limit yields $M=3\pi\mu/8 G_5$, which is an ordinary mass formula of a 
five-dimensional AdS-Schwarzschild black hole~\cite{Emp2}.

\section{Analysis}
\label{Sec:analysis}

In this and the next sections, we analyze our floating black hole initial data 
in the KR model using the Abbott-Deser mass we obtained in the previous section.
We firstly summarize the properties of the initial data in this section.
In Sec.~\ref{Sec:extrema}, we calculate the mass of the initial data for fixed values of 
entropy to study thermodynamic stability of the system in a microcanonical ensemble.
In Sec.~\ref{Sec:static}, we estimate how an initial data is close to a static solution
by observing additional matter that we have to put on the brane to make it static.

\subsection{Local extrema of entropy}
\label{Sec:extrema}

In this subsection, we calculate the Abbott-Deser mass of the initial data and 
try to find an initial data that is close to a static solution.
To be more precise, we try to find an initial data that realizes maximum entropy keeping its mass fixed.
The logic behind this search is as follows.
If we have a set of all possible time-symmetric initial data, the all static solutions
must be contained in this set since a static solution is also time-symmetric.
Since we have the area increase law of the event horizon, a time-symmetric initial data that 
realizes the maximum entropy must be a static solution.
Of course we can search only a finite region of the initial data phase space, and hence the initial data 
that realizes the maximum entropy in our initial data family 
is not necessarily static. Nevertheless, we can expect it to reflect some properties 
of a static solution since it satisfy some of the necessary conditions of a static solution.
We later compare this initial data with maximum entropy to a four-dimensional static CFT star.

As we mentioned in Sec.~\ref{Sec:ID} briefly, 
we focus on the regime that the five-dimensional model is only slightly detuned
from the RS model, i.e.~$\tk/k\ll 1$, in this paper.
In the following numerical analysis, we set $\tk/k =1/100$.
Every results are almost irrelevant to this ratio if it is smaller than this value.
For technical  convenience,
we search initial data that minimize the mass keeping the entropy fixed
instead of searching the one that maximize the entropy for fixed mass,
since the former is much simpler to implement in our initial data construction method.
This search is totally  equivalent to the search for the entropy-maximum initial data keeping the 
mass fixed.

\FIGURE[ht]{
 \centering
\includegraphics[width=12cm, clip]{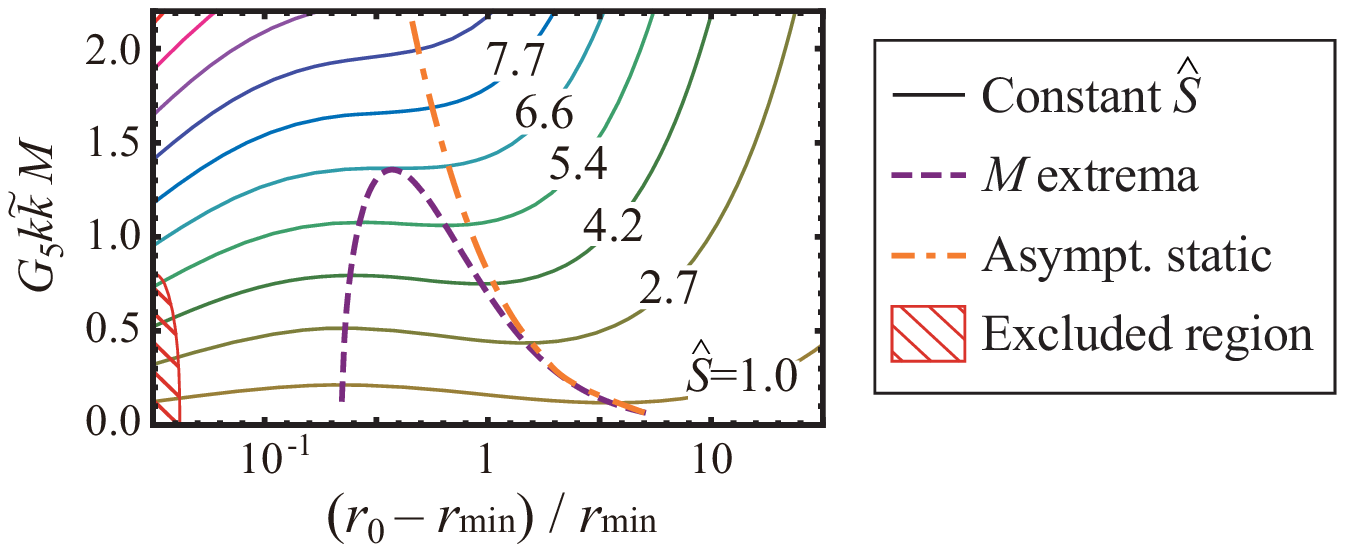}
\caption{Plot of mass $M$ for initial data sequences for fixed values of entropy 
$\Hat S\equiv G_5(k\tk)^{3/2}S$ for $\tk/k=1/100$.
The horizontal axis shows the distance between the brane position on the rotational axis $r_0$
and the lower bound for $r_0$, $r_\text{min}$, where $r_\text{min}$ takes different values on different $\Hat S$ constant sequences.
A sequence for $\Hat S$ smaller than $6.6$ possesses a local maximum and a minimum of mass.
By connecting the local extrema of various sequences, we obtain the dashed line in the plot, which have a peek at 
$G_5k\tk M=1.4$~. The dash-dotted line represents initial data that becomes static in the asymptotic region.
When the entropy is small, we see that the initial data that realizes the mass local minimum is almost 
 asymptotically static. In any initial data sequence, the mass local maximum is always further than the local 
 minimum from the point of the asymptotically-static initial data.
The initial data in the shaded region possesses an outer apparent horizon, and this region is excluded 
 from the analysis in this paper.
}
\label{Fig:sequences}
}

Figure~\ref{Fig:sequences} shows variance of the mass $M$ for sequences of fixed values of the entropy
$S= A_\text{BHs}/4G_5$.
We can see that a mass minimum and a maximum appear in each sequence for small entropy, while no 
extrema appears in a sequence for large entropy.
At $G_5(k\tk)^{3/2}S=6.6$, sequences of mass maximum and of mass minimum annihilate in pair.
The disappearance of mass local minimum suggests that 
the thermodynamical stability in a microcanonical ensemble changes from stable to unstable there,
 and that 
a transition from bulk floating black hole to the brane-localized black holes occurs near this point.

\FIGURE[ht]{
\centering
\includegraphics[width=3.0cm,clip, angle=270]{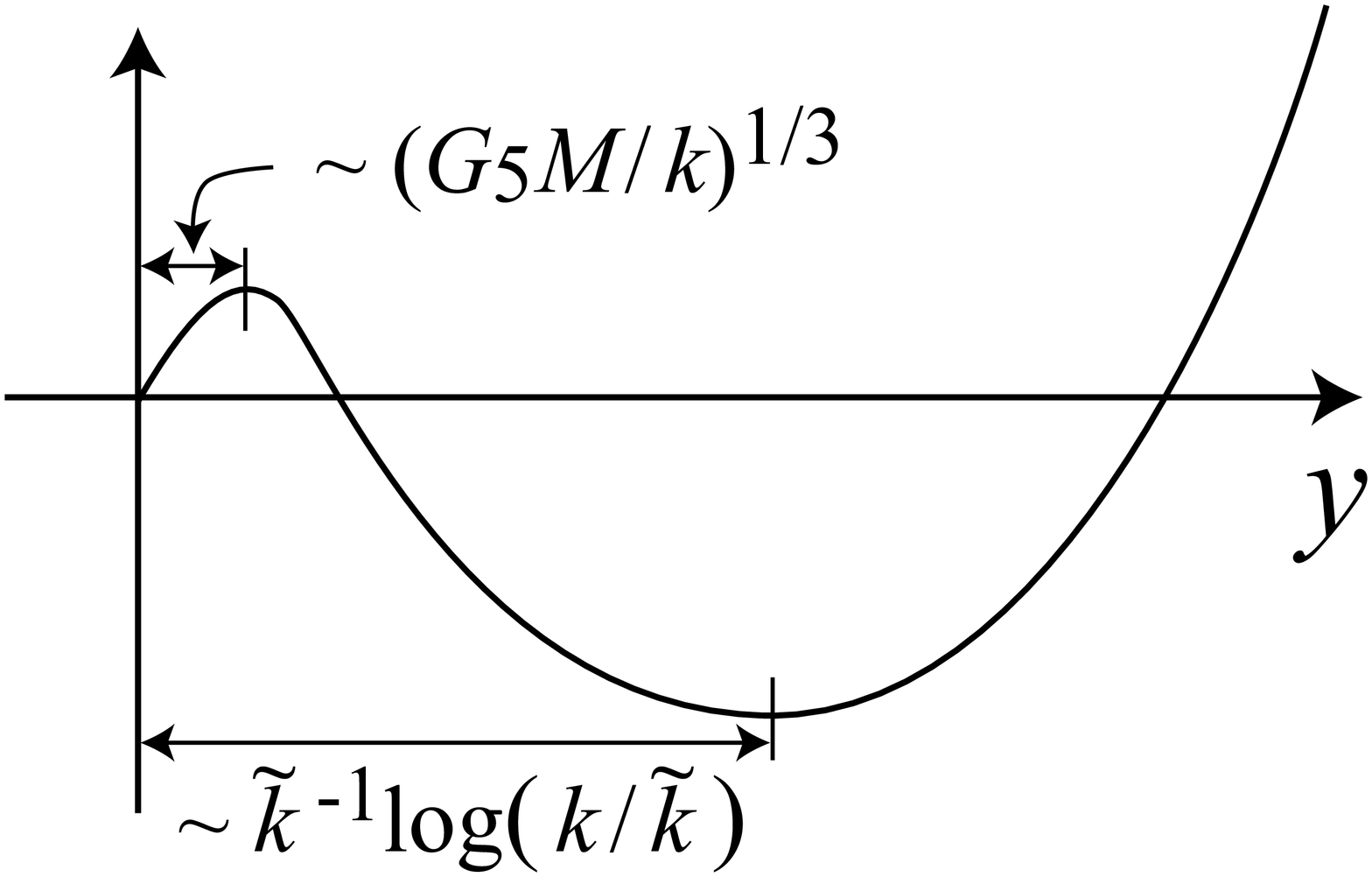}
\caption{Schematic picture of the effective potential for a small particle in the bulk of the KR model.}
\label{Fig:potential}
}

Let us further investigate the sequence of mass extrema.
The line of mass local minima is smoothly connected with the line of mass local maxima, as 
is shown in Fig.~\ref{Fig:sequences}.
We can  interpret it in the Newtonian sense as follows.
When the floating black hole is sufficiently small, it will be approximated by 
a small particle of mass $M_{sp}$. Then, there are two positions in the bulk
where this small particle can stay in a static manner: the 
potential minimum, that is the throat of the AdS$_5$ at which the warp factor is minimized, 
and the near brane point at which the brane's 
repulsion force $a=k$ is cancelled by the gravitational attraction due to its mirror image behind 
the brane $a\sim -G_5M_{sp}/(2d)^3$, where $d$ is proper distance between the small particle and the brane.
We show a schematic picture of the effective gravitational potential for a small particle 
in Fig.~\ref{Fig:potential}.
See also \cite{TanakaFloat,Kashiyama} for detail about this effective potential. 
When the small particle sits at the latter point near the brane,
the system is in unstable equilibrium and the energy will be 
maximized.
For example, mass-maximum initial data in the sequence of $G_5 (k\tk)^{3/2}S=2.0\times 10^{-3}$ 
has black holes with $G_5 k\tk M_{sp}\sim G_5 k\tk M/2=1.3\times 10^{-2}$,
and then the position of the unstable equilibrium is estimated as 
$\tk d=\tk(G_5M_{sp}/8k)^{1/3}=5.4\times 10^{-3}$.
This value is roughly equal to 
the actual proper distance between the brane and the black hole in our initial data, which is $4.4\times 10^{-3}$.
As the floating black hole becomes large, those two points of equilibrium get closer to each other 
and finally annihilate. This annihilation corresponds to the turn over of the mass-extrema sequence 
in Fig.~\ref{Fig:sequences}.

We find some initial data, especially those with small floating black hole located very near the 
brane, have an outer apparent horizon. We exclude such initial data in this paper to focus only 
on the bulk floating black hole solutions. 
We also mention that any initial data with maximum or minimum mass seems not to have an outer apparent 
horizon, at least for $G_5 (k\tk)^{3/2}S \gtrsim 1$.

\subsection{Difference of initial data from static solutions}
\label{Sec:static}

In our initial data construction, the brane does not become static in the GN coordinates 
by itself in general.
Since we would like to compare the initial data with static CFT stars, we have to know how ``far'' 
an initial data is  from a static state.
For this purpose, we try to find asymptotically-static initial data, in which 
the brane becomes static in the asymptotic region.

Since the transformation (\ref{trsf}) 
maps a static brane in the GN coordinates to a brane with $\chi_0=\pi/2$ in the AdS global coordinates
and 
it does not mix the temporal coordinate with the spatial coordinates,
a brane with $\chi_0=\pi/2$ becomes static at $\la\to\infty$.
We denote an initial data with such a brane as an asymptotically-static initial data.
We can find such an initial data uniquely in an initial data sequence for a fixed $S$.
The dash-dotted line in Fig.~\ref{Fig:sequences} is of the sequence of such asymptotically-static initial data.
In the figure, we find that the mass-minimum initial data are approximately asymptotically static when 
the entropy is small.
We can also see that
 the mass-maximum initial data is always more distant from the asymptotically static one
than the mass-minimum one with the same entropy.

The result that a mass-minimum initial data becomes closer to an asymptotically-static initial data 
for small bulk black holes is a natural one in the 
following sense; When the bulk black hole is small, it will be approximated by a small particle. If 
we place it at the potential minimum of the bulk gravitational field, it will stay there in a static 
manner without disturbing the background geometry. Then, the whole system including the small 
particle will be almost static, and it approximates the asymptotically-static initial data well.

We can assess the staticity of an initial data in a more quantitative way as follows.
When a brane in an initial data is dynamical, we can make it static by putting some additional matter 
on it. The amount of this additional matter will be a good indicator of staticity.
By comparing it with the effective energy-momentum tensor or the 
cosmological constant on the brane, we can see to what extent the initial data is close to a static solution.

The additional matter to make the brane static is estimated as follows.
The four-dimensional Einstein equations on the brane are given by
\begin{equation}
 G^\text{4D}_{ab}+\Lambda_{4}g^\text{4D}_{ab}=-\mathcal{E}_{ab}~~,
\label{G4D}
\end{equation}
where $\mathcal{E}_{ab}\equiv C_{\m\la\n\ro}\gamma^\m_{~a} s^\la \gamma^\n_{~b} s^\ro $ is the bulk 
Weyl tensor projected onto the brane and $G^\text{4D}_{ab}$ is the Einstein tensor for four-dimensional 
metric induced on the brane~\cite{SMS}.
Effective energy density and pressure on the brane are defined 
using four-dimensional Newton constant $G_4=kG_5$ 
as $\rho\equiv \mathcal{E}^\tau_{~\tau}/8\pi G_4$ 
and $P_i\equiv -\mathcal{E}^i_{~i}/8\pi G_4$ for $i=\la$ and $\theta$, respectively.
To keep the brane static, we have to add
an additional energy-momentum tensor $\delta T_{ab}$, 
which replaces Eq.~(\ref{G4D}) with
\begin{equation}
G^\text{4D}_{ab}+\Lambda_{4}g^\text{4D}_{ab}=-\mathcal{E}_{ab}+8\pi G_4 \delta T_{ab}~~.
\label{G4Dadd}
\end{equation}
Now that the brane is static, the induced metric on the brane is given by
\begin{align}
 ds^2=g^\text{4D}_{ab}dx^adx^b&=
-U dt^2
+\left(\frac{1}{U}+r^2\sin^2\chi\right)dr^2
+r^2\sin^2\chi \left( d\theta^2+\sin^2\theta d\phi^2  \right)
\no
&=
-\biggl(\frac{\tk}{k}\biggr)^2 Ud\tau^2 
+\frac{1}{\tU'}d\la^2 +\la^2 
\left( d\theta^2+\sin^2\theta d\phi^2  \right)~~,
\label{induced}
\end{align} 
where $\tU'$ is defined by the last equality of this equation.
Using this metric, $\delta T_{ab}$ is obtained via Eq.~(\ref{G4Dadd}).
By comparing $\delta T_{\m\n}$ with $-\mathcal{E}_{ab}/8\pi G_4$ or $\Lambda_\text{4}$, we can 
estimate how the brane is close to a static state.
We note here that $\delta \rho\equiv -\delta T^\tau_{~\tau}$ automatically vanishes since the Hamiltonian 
constraint is satisfied. Below, we investigate properties of 
the nontrivial components $\delta P_i\equiv \delta T^i_{~i}$.


\FIGURE[ht]{
 \centering
\includegraphics[width=15.1cm, clip]{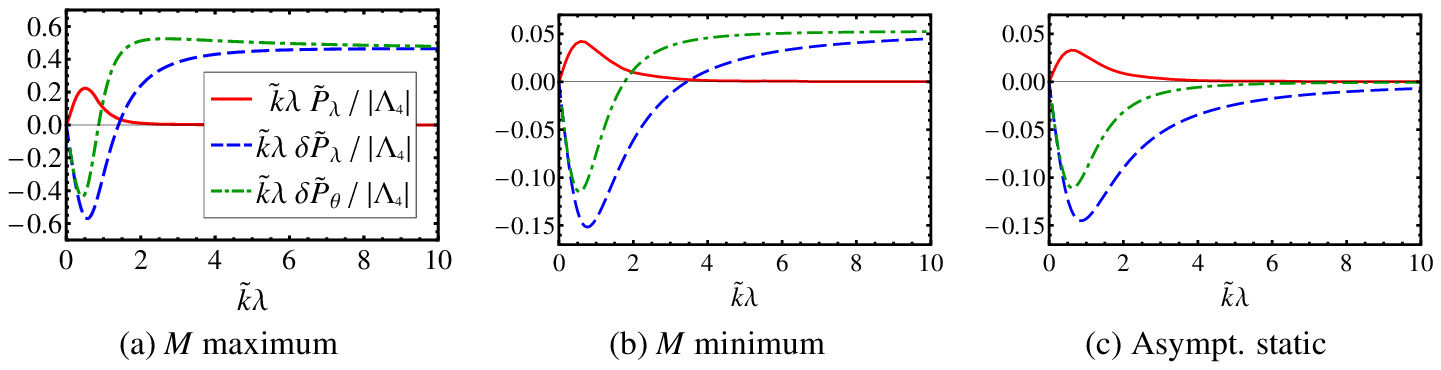}
\caption{
Distributions of effective pressure $\tilde P_\la\equiv 8\pi G_4 P_\la$ and additional
pressure $\delta \tilde P_i \equiv 8\pi G_4 P_i$ for $i=\la$ and $\theta$.
The panels (a) and (b) show the distributions for initial data with maximum and minimum $M$ 
of an initial data sequence for $G_5(k\tk)^{3/2}S=4.3$~, and the panel (c) shows that for the 
 asymptotically-static initial data in the same sequence.
The all quantities are normalized by $\Lambda_4$.
$\delta P_i$ decays as $\la^{-1}$ and 
is much smaller for the asymptotically-static initial data than the others. 
Note that scale of the vertical axis in the panel (a) is much larger than those of the other panels.
}
\label{Fig:delta}
}

In Fig.~\ref{Fig:delta},
we plot distributions of effective matter and additional matter 
for mass-maximum, mass-minimum and asymptotically-static initial data in the initial data sequence 
for $G_5(k\tk)^{3/2}S=4.3$~.
These plots shows that additional matter of much larger amount is necessary to make the mass-maximum 
initial data static compared to the mass-minimum and asymptotically-static ones.
Note that scale of the vertical axis of Fig.~\ref{Fig:delta}(a) is much larger than those of 
Figs.~\ref{Fig:delta}(b) and \ref{Fig:delta}(c).
This confirms that the mass-minimum initial data is closer to a static solution compared to the 
mass-maximum one.

This fact about mass-maximum initial data 
indicates that the genuine static solution of the floating black hole in unstable 
equilibrium, even if existed,
 is far from the initial data we constructed.
The reason for it may be as follows: the floating black hole in the unstable equilibrium is 
highly deformed by the self-gravity due to its mirror image behind the brane and the repulsion force 
due to the brane, and then it cannot be 
well approximated by an AdS-Schwarzschild black hole we used in the initial data construction.

\section{Comparison with Four-dimensional CFT stars}
\label{Sec:CFTstar}

Using the tools prepared in the previous sections, we conduct a comparison between the 
five-dimensional floating black hole initial data and the four-dimensional CFT stars.

We firstly illustrate the numerical construction method of the CFT star configuration
and summarize basic properties of the CFT stars in Sec.~\ref{Sec:CFTstarConstruction}, 
though the construction method is originally 
 given by~\cite{PP} and a detailed analysis is done by~\cite{Kashiyama}.
In Secs.~\ref{Sec:sequence} and \ref{Sec:critical}, we compare the sequence of the CFT stars with 
the sequence of initial data at mass extrema, which is expected to mimic a sequence of 
five-dimensional static solutions.
We compare the structure of the whole sequences in Sec.~\ref{Sec:sequence}, and after that we 
focus on some specific solutions in the sequences in Sec.~\ref{Sec:critical}.

\subsection{Construction method}
\label{Sec:CFTstarConstruction}

When the temperature is sufficiently high compared to the characteristic energy scale on the brane%
\footnote{
When the temperature measured on the brane, $T$, is higher than $\mathcal{O}\bigl(\tk\bigr)$, the bulk 
black hole temperature in the global coordinates is higher than $\mathcal{O}(k)$. 
It implies that the five-dimensional bulk is in the AdS-Schwalzschild black hole phase not in 
the thermal AdS phase for this high-temperature regime.
We focus only on this regime in this paper.
},
$T\gg \tk$, we can approximate the CFT with radiation 
fluid, whose energy density is given by $\rho=aT_\text{loc}^{~~4}$ where $T_\text{loc}$ is the local 
temperature determined from the global temperature $T$ as $T_\text{loc}=(-g_{\tau\tau})^{-1/2}T$.
In our setting, two sets of the CFT reside on the brane since we have two bulk regions on each side 
of the brane. Then, the radiation density constant is given by $a=(\pi^2/30)\times 15N^2\times 
2=\pi^3/2k^3G_5$, where the factor $15=6+2+(7/8)\times 8 $ is the effective spin state number of the 
CFT multiplets:  six real scalars, one vector and four Weyl fermions.

Assuming spherical symmetry and staticity, general four-dimensional metric can be parametrized as
\begin{equation}
 ds^2=-e^{2\psi}V(\la)dt^2+V(\la)^{-1}d\la^2
+\la^2\left( d\theta^2 + \sin^2\theta d\phi^2 \right)
\end{equation}
and $V=1+\tk^2\la^2-2m(\la)/\la$.
We require that $\lim_{\la\to\infty}\psi=0$ to realize asymptotically AdS$_4$ spacetime
and $m(\la=0)=0$ to maintain the regularity at the center.
Then, the four-dimensional Einstein equations for energy-momentum tensor of the radiation fluid 
$T^\m_{~\n}=\text{Diag}[-\ro,p,p,p]$ with $\ro=3p=aT^4e^{-4\psi}V^{-2}$ are given by
\begin{gather}
 \frac{dm}{d\la}=4\pi \la^2 \ro~, 
\qquad
\frac{d\ro}{d\la}=\frac{-4\ro\left(G_4m+4\pi G_4\la^3\ro/3+\tk^2\la^3\right)}
{\la^2+\tk^2\la^4-2G_4\la m}~.
\end{gather}
Assuming regularity at $\la=0$ and giving central density $\ro(\la=0)\equiv \ro_c$, 
we can solve these equations to 
obtain one-parameter family of solutions $m(\ro_c;\la)$ and $\ro(\ro_c;\la)$.
Four-dimensional mass is given by $M=\lim_{\la\to\infty}m(\la)$.
Using the solutions $m$ and $\ro$,
the global temperature and the metric function $\psi$ 
can be obtained from the expression of the energy density 
as $T=\lim_{\la\to\infty}(\ro V^2/a  )^{1/4}$ and $\psi=(1/4)\log(aT^4/\ro V^2)$.
The total entropy of the system $S$ can be obtained by integrating 
proper entropy density $s=(4/3)aT_\text{loc}^{~~3}$.
We note this $S$ satisfies the thermodynamic first law $dM=TdS$.

Since the CFT resides not only on the brane but also on the AdS boundaries, we have to add their
contributions to $M$ and $S$.
For convenience of the derivation, 
let us consider from the five-dimensional point of view.
We introduce a second AdS brane of curvature $\tk_b$ 
in the bulk near the AdS boundary, and consider the CFT on it.
Let us refer to this CFT on the second brane as CFT2.
In the GN coordinates~(\ref{GN}), such a brane is at $y=y_b$ such that 
$(\tk/k)\cosh(ky_b)=\tk/\tk_b$.
The induced geometry on that second brane is given by
\begin{equation}
 ds^2=
\biggl(\frac{\tk}{\tk_b}\biggr)^2
\left(
-\tU d\tau^2+\frac{d\la^2}{\tU}+\la^2 d\Omega_\text{II}
\right)
=-\tU_b d\tau_b^2 +\frac{d\la_b^2}{\tU_b}+\la_b^2 d\Omega_\text{II}~~,
\label{second}
\end{equation}
where $\tU=1+\tk^2\la^2$ and $\tU_b=1+\tk_b^2\la_b^2$.
From this equation, we find that the coordinates on the brane $\tau_b$ and $r_b$ are related to those 
in the GN coordinates as 
\begin{equation}
\tau_b/\tau=\la_b/\la=\tk/\tk_b~~.
\label{rescale}
\end{equation}
The CFT on the second brane have to be in thermal equilibrium with that on the first brane.
Hence, the 
global temperature on the second brane, which is the equilibrium temperature measured by $\tau_b$, 
is given by $T_b=(\tau/\tau_b)T=(\tilde k_b/\tk)T$.
We note here that the radiation fluid approximation is valid also on the second brane 
since this relation and the condition $T\gg\tk$ implies $T_b\gg\tilde k_b$.
Since the CFT decouples from gravity in the limit to take the second brane to the AdS boundary,
we can neglect the back reaction of the radiation to the background geometry.
Then, energy density and entropy density are given by $\rho=aT^{~~4}_\text{loc}$ and $s=(4/3)aT_\text{loc}^{~~3}$
with $T_\text{loc}=\tU_b^{-1/2}T_b$, 
and total mass and entropy are given by
\begin{equation}
M_b = \int_0^\infty 4\pi \la_b^2\rho \,d\la_b = \frac{\pi^2a}{\tk^4}\tk_b T^4~~,
\qquad
S_b = \int_0^\infty 4\pi \la_b^2 s \;\sqrt{g_{\la_b\la_b}}d\la_b= \frac{4\pi^2a}{3\tk^3} T^3~~.
\end{equation}
Due to the rescaling of the time coordinate~(\ref{rescale}), an observer on the first brane measures 
the mass and entropy on the second brane as $M_\text{CFT2}=(\tk/\tk_b)M_b$ and 
$S_\text{CFT2} = S_b$. These $M_\text{CFT2}$ and $S_\text{CFT2}$ are independent of $\tk_b$ and of the 
brane position, we can safely take the limit to send the second brane to the AdS boundary 
making $\tk_b\to 0$.
Note that there are two AdS boundaries in our setting, and 
their contributions can be taken into account simply by 
setting the radiation density constant $a$ to $\pi^3/2k^3G_5$ in the above calculation.
Below in this paper, we always include these $M_\text{CFT2}$ and $S_\text{CFT2}$ to the total mass $M$ and 
entropy $S$.

By the way, the radiation fluid approximates a weak-coupling CFT, though the five-dimensional gravity
is expected to be dual to a strongly-coupled CFT.
It is known that this gap can be filled by reducing the CFT's degrees of freedom $N^2=\pi/2k^2G_4$ 
to $(3/4)\times N^2$.
Inclusion of this factor can be easily performed as follows.
When we change $G_4$ and $k$,
$M$, $S$, $T$, $\rho$, and $G_5$ are rescaled as  
$M\propto G_4^{-1}$, $\rho\propto G_4^{-1}$, $T\propto k^{1/2}$, and $S\propto G_4^{-1}k^{-1/2}$, 
respectively.
Using this scaling, we find that,
for example, each quantity 
is rescaled as $M\to M$, $\ro\to\ro$, $T\to (3/4)^{-1/4} T$, and $S\to(3/4)^{1/4}S$
if we fix $G_4$ to be constant and introduce the factor $3/4$ by changing $k$.
We always take this factor into account in the analysis of CFT stars hereafter in this paper.

Let us summarize features of the CFT star solution sequence.
When $\rho_c$ is lower than the four-dimensional cosmological constant $\Lambda_\text{4}=-3\tk^2$, 
the background geometry does not change much from the pure AdS$_4$, and the total mass and entropy 
are proportional to $\rho_c$ and $\rho_c^{3/4}$, respectively.
In this regime, the CFT stars have positive specific heat and thus they are thermodynamically stable.
Increasing $\rho_c$, the total mass and the entropy is maximized for the same critical $\rho_c$.
The system becomes thermodynamically unstable for $\ro_c$ larger than this critical value 
if the system is in a microcanonical ensemble.

\FIGURE[ht]{
 \centering
\includegraphics[width=15.1cm, clip]{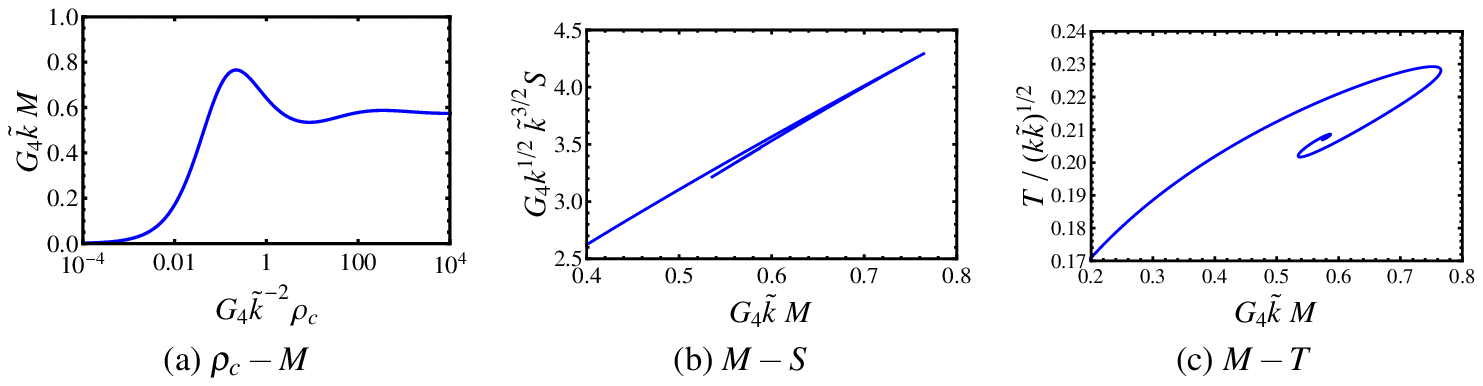}
\caption{
Plots of characteristic quantities of the CFT star sequence.
The panel (a) shows the behavior of $M$ against $\rho_c$. 
$M$ is maximized at $G_4\rho_c/\tk^2=0.22$, and it shows oscillatory behavior for larger $\rho_c$.
The system becomes thermodynamically unstable for $\rho_c$ larger than this critical value if the 
 system is in a microcanonical ensemble.
The panels (b) and (c) show variance of $S$ and $T$ against $M$, respectively.
In the both panels, the curves turn around at the critical $\rho_c$ and converges to a point in the limit of 
 infinite $\rho_c$.
}
\label{Fig:CFTstar}
}

In the limit of infinite central density, the radiation fluid may gravitationally collapse and 
form a small black hole.
In the AdS spacetime, 
this black hole can settle into thermodynamical equilibrium with its own Hawking radiation,
since the gravitational potential due to the negative cosmological constant works as an effective cavity.
We call this system in thermal equilibrium as a quantum-corrected black hole.
It is shown that the CFT star sequence is smoothly connected to the quantum-corrected black hole 
sequence in the limit of infinite central density~\cite{Kashiyama}.
This quantum-corrected black holes are conjectured to be dual to five-dimensional 
brane-localized black holes, though we do not pursue this duality between them in this paper.

\subsection{Comparison of the solution sequences}
\label{Sec:sequence}

In this and the next sections, we juxtapose the CFT star sequence with the
sequence of the five-dimensional floating black hole initial data at local extrema of mass.
Firstly, we compare the whole solution sequences of the four-dimensional CFT picture and the 
five-dimensional KR picture.
In Fig.~\ref{Fig:MST}, we show the relation of $M$, $S$ and $T$ along the solution sequences.
The curves for the CFT star sequence are the same as those in Figs.~\ref{Fig:CFTstar}(b) 
and \ref{Fig:CFTstar}(c).
For initial data, the temperature $T$ is not well-defined since the 
information of the lapse function, which is crucial for the temperature determination, is missing.
We thus define $T$ of initial data using the thermodynamic first law $T\equiv dM/dS$.
This $T$ will approximate the physical temperature of the system if the initial data is close to a 
static solution.

\FIGURE[ht]{
 \centering
\includegraphics[width=14cm, clip]{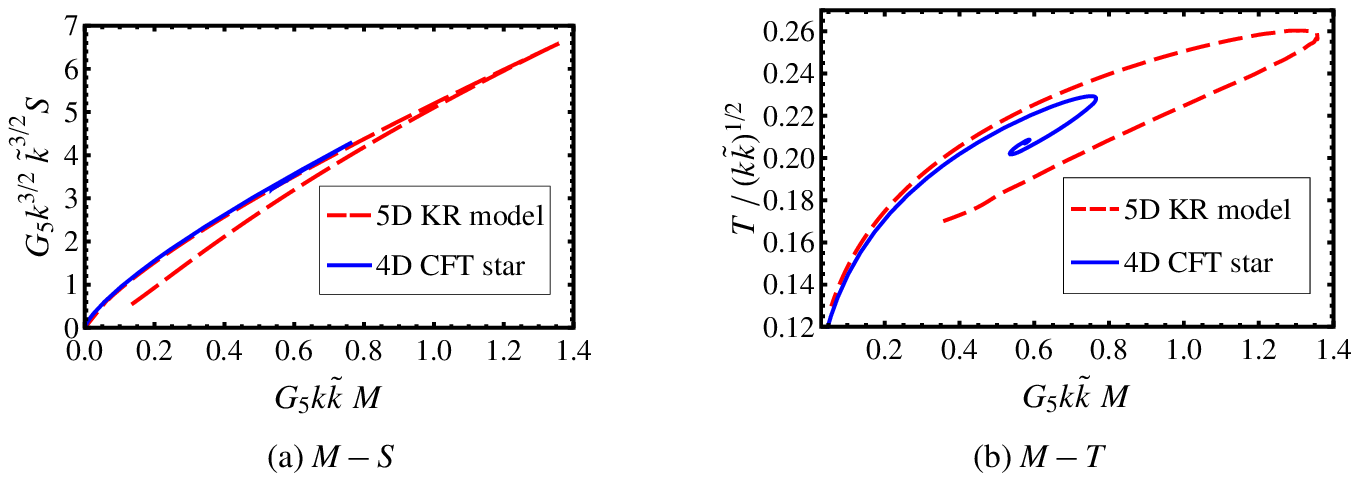}
\caption{
$M-S$ and $M-T$ diagram for solution sequences.
The dashed line is of the five-dimensional initial data at the mass extrema, and the solid line is 
 of the four-dimensional CFT stars.
In the left panel for $M-S$ relation,
the curves of the two pictures have sharp edges at certain critical values of mass, which are $G_5k\tk M=0.77$ for the 
 four-dimensional CFT star sequence and $1.4$ for the five-dimensional initial data sequence.
We can see qualitative similarities between the curves of the two pictures in each panel, while
 there are quantitative differences.
The curves of the two pictures match each other well in the range 
from the origin to the critical $M$ of the CFT star sequence. 
The deviation of the two curves becomes significant 
in the parameter region where the initial data become far from static solutions 
(see Fig.~\ref{Fig:sequences}).
}
\label{Fig:MST}
}

In the plot, we can see some common features between the curves of the initial data and of the CFT stars.
The curves near the origin of the graph correspond to the sequences of small floating black holes or 
low central density CFT stars.
As we increase the black hole size or the central density, they approach the 
critical points at which $M$ is maximized, and then they turn around.
In this sense, thermodynamic phase structures are qualitatively the same between the two pictures.
Since 
the Hawking-Page transition occurs in the four-dimensional picture and a transition of the bulk 
black hole configuration is suggested to occur (see Sec.~\ref{Sec:extrema}) at the critical points, 
this result suggests that these transitions in the two pictures
are holographic dual of each other.

A quantitative difference appears in the part of the sequences after the critical point
of the CFT star sequence, which is at
$(G_5k\tk M, G_5 (k\tk)^{3/2}S, T/(k\tk)^{1/2})=(0.77, 4.3, 0.23)$~.
No feature appears in the initial data sequence at the critical point of the CFT star sequence, 
and the critical point in the five-dimensional initial data sequence appears at 
$(G_5k\tk M, G_5 (k\tk)^{3/2}S, T/(k\tk)^{1/2})=(1.4, 6.6,0.26)$~.
A conceivable reason for this mismatch is that the initial data in that parameter region is far from 
a static state. 
Figures~\ref{Fig:sequences} and \ref{Fig:delta} show that mass-maximum initial data are
very different from static solutions, and deviation of an initial data from a static solution 
becomes larger as we move along the sequence toward the mass-maximum initial data.
When the staticity is largely violated,
the duality between 
the mass-extremum initial data and the static CFT stars will break down.
In fact, 
the deviation of the initial data from asymptotically-static solutions becomes manifest for 
$G_5 (k\tk)^{3/2}S\gtrsim 4$ as is shown in Fig.~\ref{Fig:sequences}.
This naive expectation will be confirmed using five-dimensional static solutions, which are yet to 
be constructed.

\subsection{Comparison of solutions}
\label{Sec:critical}

Next, we compare some specific solutions in the both pictures to
investigate to what extent the duality holds.
In the CFT picture, we take a CFT star solution at the critical point, at which $M$ is maximized,
as a representative.
In the KR picture, we choose a mass-minimum initial data 
which share the same $M$ with that CFT star solution.
In other words, we pick up solutions 
near the point $(G_5k\tk M, G_5 (k\tk)^{3/2}S)=(0.77, 4.3)$ in Fig.~\ref{Fig:MST}(a)
from both sequences, and compare those solutions to each other.
%
We compare the energy-momentum tensor of the four-dimensional CFT star $T_{ab}$ with the effective 
energy-momentum tensor $-\mathcal{E}_{ab}/8\pi G_5$
between such solutions.
This effective energy-momentum tensor is traceless by definition, which is a common property with $T_{ab}$
 of the radiation fluid.
We would like to clarify other similarities between them in this subsection.
Adding to the energy-momentum,
we also compare the four-dimensional geometries between the two pictures.

\FIGURE[ht]{
  \centering
\includegraphics[width=14cm, clip]{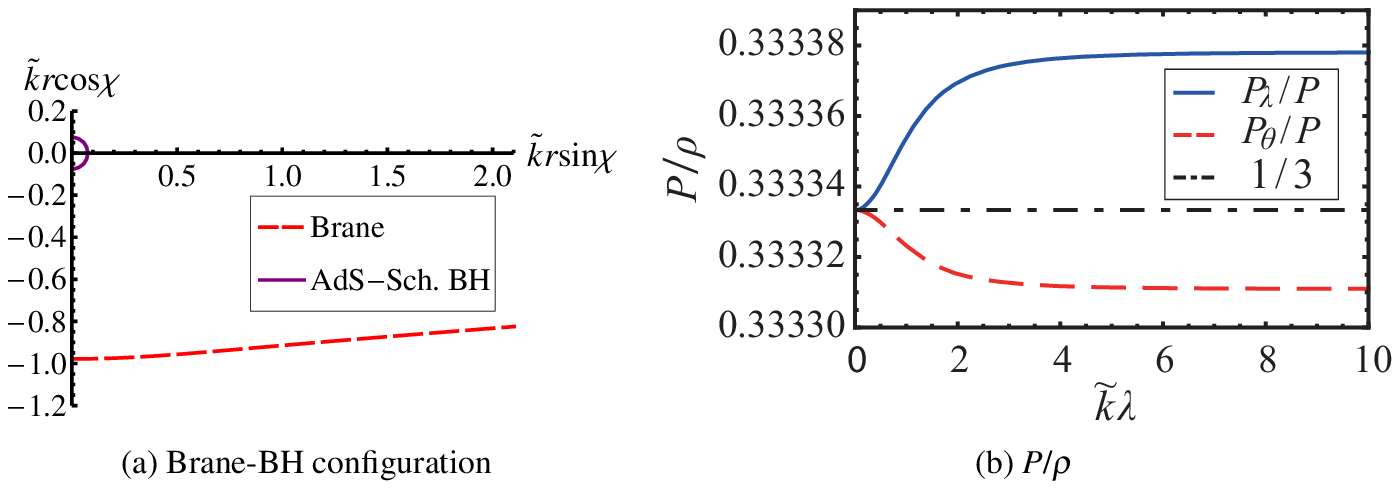}
\caption{
Initial data at the data point $(G_5k\tk M, G_5 (k\tk)^{3/2}S)=(0.77, 4.3)$
and the effective pressure on its brane. 
The left panel shows the configurations of the brane and the 
 bulk floating black hole. The bulk region is the upper side of the brane, 
and the  lower side of the brane is to be cut away.
The right panel shows the ratios of the effective pressure components $P_i$ in radial and angular directions
to the effective energy density $\rho$ on the brane.
Relative difference of those ratios from $1/3$ is smaller than a percent.
It shows that the effective pressure on the brane is almost isotropic.
}
\label{Fig:5D}
}

In Fig.~\ref{Fig:5D}(a), we show the configuration of the brane and the floating black hole
for the data point mentioned above.
As a first check, 
we study properties of the effective pressure distribution in Fig.~\ref{Fig:5D}(b).
We find that the pressure is approximately isotropic, i.e.~$P_\la\simeq  P_\theta$, similarly to the
radiation fluid.
Since the brane's effective energy-momentum tensor is traceless, i.e.~$\rho=P_\la+2P_\theta$, this 
result means that the equation of state of the effective matter is quite similar to the radiation fluid.

\FIGURE[ht]{
 \centering
\includegraphics[width=14cm, clip]{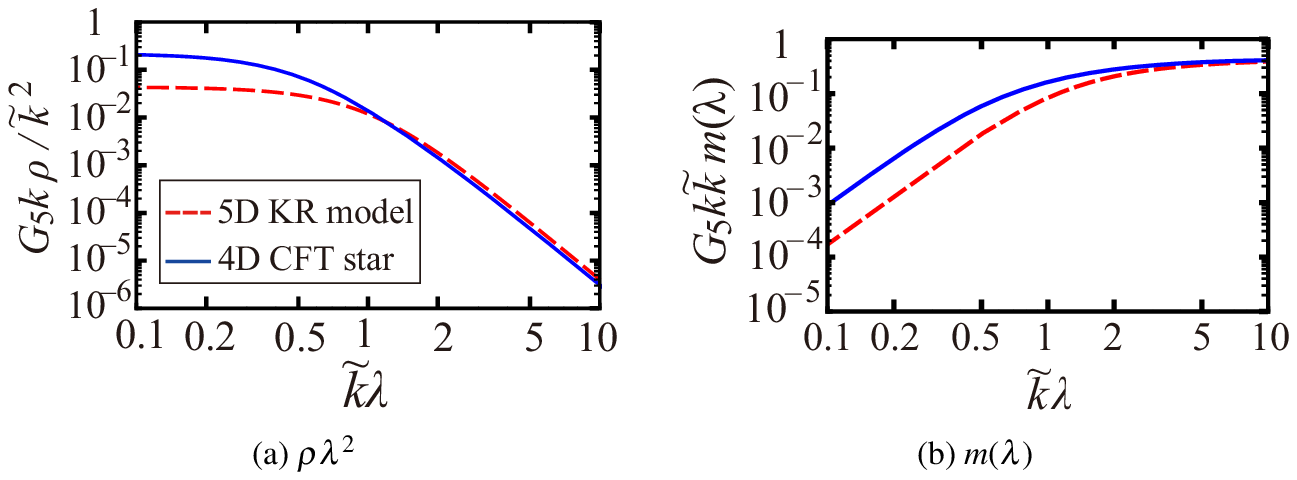}
\caption{
Comparison of characteristic quantities 
at the data point $(G_5k\tk M, G_5 (k\tk)^{3/2}S)=(0.77, 4.3)$~.
The left panel shows distributions of effective energy density on the brane and of the CFT star. 
The right panel shows the metric function $m(\la)$ in $\Tilde U=1+\tk^2 \la^2-2G_4m(\la)/\la$.
In both figures, we can see that the curves of the two pictures are close for a large $\la$, while they 
 deviate near the center $\la=0$.
}
\label{Fig:comparison}
}

\FIGURE[ht]{
 \centering
\includegraphics[width=14cm, clip]{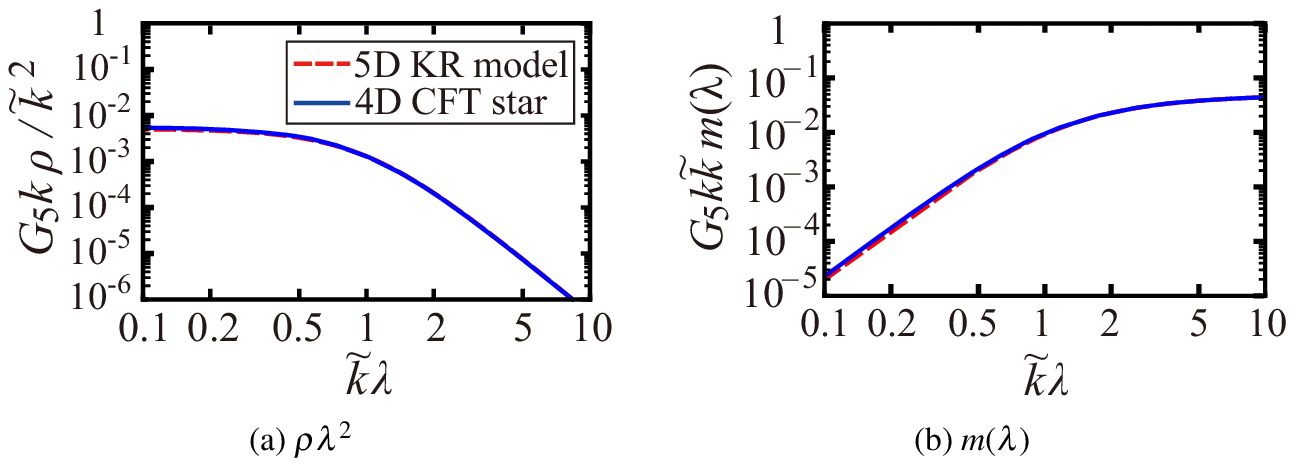}
\caption{
Comparison of characteristic quantities 
at the data point $(G_5k\tk M, G_5 (k\tk)^{3/2}S)=(0.10, 0.89)$~,
at which the initial data is almost asymptotically static as we can see in Fig.~\ref{Fig:sequences}.
In this case, almost exact match is realized between $\rho$ or $m(\la)$ of the two pictures.
}
\label{Fig:smallM}
}

In Fig.~\ref{Fig:comparison}, we compare radial profiles of energy density $\rho$ and the metric function $m(\la)$.
Firstly, we compare $\rho$ of the two pictures in Fig.~\ref{Fig:comparison}(a).
We can see that the curves of the two pictures share common qualitative behaviors, such as power-law 
fall-off in the asymptotic region $\la\to\infty$.
Secondary, Fig.~\ref{Fig:comparison}(b) shows that 
the intrinsic geometry of the four-dimensional spacetime in the CFT picture 
is quite similar to that induced on the brane in the five-dimensional KR picture.
Asymptotic values of $m(\la)$ in the two pictures are almost same%
\footnote{
Since four-dimensional graviton is massive in the KR model,
we expect that $m(\la)$ decays to zero in the asymptotic region for a static solution. 
This decaying property 
of $m(\la)$ should appear in the four-dimensional CFT picture if we take into account the screening 
effect of mass due to CFT correction to the graviton~\cite{P,Duff}, 
while this effect is neglected in the approximations adopted in this work.
It will be interesting to pursue this issue using five-dimensional static solutions and 
four-dimensional solutions with CFT-dressed graviton.
}%
, which is a non-trivial result since we 
tuned the total mass $M$ of the solutions to be common but did not tuned the value of 
$\lim_{\lambda\to\infty}m(\la)$ directly.
Recall that $M$ of a CFT star is given by a sum of the asymptotic value of $m(\la)$ and the mass of 
the CFT2 as we explained in Sec.~\ref{Sec:CFTstarConstruction},
and that $M$ of an initial data is not related directly to $m(\la)$ 
determined from the induced geometry on the first brane.
These similarities between the two pictures supports the conjectured correspondence in the KR model.
However, quantitative difference between the curves becomes big in the central region $\la\simeq 0$
as we can see in the both plots. 
We expect this discrepancy to be reduced if we use static solutions instead of initial data for comparison.

In Fig.~\ref{Fig:smallM}, we show $\rho$ and $m(\la)$ for the data point 
$(G_5k\tk M, G_5 (k\tk)^{3/2}S)=(0.10, 0.89)$, for which the mass-minimum initial data 
is almost equivalent to an 
asymptotically-static initial data (see Fig.~\ref{Fig:sequences}).
In this case, 
two pictures give almost exactly the same $\rho(\la)$ and $m(\lambda)$. 
This fact supports the above naive expectation that 
the coincidence will improve  for five-dimensional static solutions.

\section{Summary}
\label{Sec:summary}

Using the time-symmetric initial data,
we showed that the Hawking-Page transition in a microcanonical ensemble
on four-dimensional spacetime is well reproduced 
in the five-dimensional KR model as a transition of 
bulk black hole configuration.
We found good 
 agreements between those two pictures
in thermodynamical phase structures 
and in some characteristic values such as 
energy density distributions or the intrinsic geometries.
These results strongly suggest that the holographic description initially claimed in the RS model
also holds in the KR model.

As a byproduct, 
we obtained a definition of conserved mass in the KR model by extending the definition 
of mass for asymptotically AdS 
spacetime by Abbott and Deser.
We also gave an explicit mass formula for the initial data we constructed.
This definition reproduces an ordinary mass formula for an AdS-Schwarzshilcd black hole.
Since this definition of mass is valid even for dynamical spacetime,
it will have a wide application to the analysis of gravitational phenomena in the KR model.

Although we used time-symmetric initial data in this study,
it is definitely preferred to use static solutions to compare with static CFT stars, which will 
improve the agreement between the two pictures significantly. Such five-dimensional static solutions 
may be constructed only numerically. We will study this issue in our future work.
In addition to that, 
it was conjectured in Ref.~\cite{Tanaka}
that brane-localized black holes in the KR model are dual to four-dimensional quantum-corrected black holes, 
whose properties are 
intensively studied in Ref.~\cite{Kashiyama}.
Investigation of this duality using five-dimensional initial data or static solutions
will clarify further the properties of the holography in the KR model.

Another ambitious future work 
is to test the holography of this type in dynamical set-up using the technique of
numerical relativity 
in higher-dimensional spacetime, which is now available in asymptotically flat spacetime~\cite{YS,Abe}. 
The initial data studied in this paper also provide a steady step toward this project.
We would like to pursue it in the future.

\acknowledgments

This work is in part supported by
Grant-in-Aid for Scientific Research Nos.~2056381,
19540285 and 21244033.
This work is also supported by the Grant-in-Aid for the Global COE Program 
"The Next Generation of Physics, Spun from Universality and Emergence" 
from the Ministry of Education, Culture, Sports, Science and Technology (MEXT) of Japan.
The authors
thank Kazumi Kashiyama and Antonino Flachi.

%

\end{document}